\documentclass[final,twocolumn,english,prl,notitlepage,nofootinbib,floatfix,longbibliography,superscriptaddress
]{revtex4-2}
\usepackage[T1]{fontenc}
\usepackage[utf8]{inputenc}
\usepackage{refstyle}
\usepackage{amsmath}
\usepackage{bm} 
\usepackage{amssymb}
\usepackage{graphicx}
\usepackage[pdftex, hidelinks]{hyperref}
\usepackage[usenames,dvipsnames]{xcolor}
\usepackage{bbm}
\usepackage{tikz}
\usepackage{xspace}
\usepackage{bm}
\usepackage{booktabs}
\usepackage{babel}
\usepackage[separate-uncertainty=true]{siunitx}
\usepackage[makeroom]{cancel}
\usepackage[obeyFinal,colorinlistoftodos,color=green]{todonotes} 
\usepackage{ifdraft} 
\usepackage[normalem]{ulem} 

\usetikzlibrary{matrix}
\newcommand\myshade{85}
\colorlet{mylinkcolor}{violet}
\colorlet{mycitecolor}{YellowOrange}
\colorlet{myurlcolor}{Aquamarine}
\hypersetup{
  linkcolor  = mylinkcolor!\myshade!black,
  citecolor  = mycitecolor!\myshade!black,
  urlcolor   = myurlcolor!\myshade!black,
  colorlinks = true,
}

\DeclareMathAlphabet\mathbfcal{OMS}{cmsy}{b}{n}

\makeatletter

\def\@fnsymbol#1{\ensuremath{\ifcase#1\or *\or *,\dagger\or \ddagger\or
   \mathsection\or \mathparagraph\or \|\or *\or \dagger\dagger
   \or \ddagger\ddagger \else\@ctrerr\fi}}

\AtBeginDocument{}
\RS@ifundefined{subsecref}
  {\newref{subsec}{name = \RSsectxt}}
  {}
\RS@ifundefined{thmref}
  {\def\RSthmtxt{theorem~}\newref{thm}{name = \RSthmtxt}}
  {}
\RS@ifundefined{lemref}
  {\def\RSlemtxt{lemma~}\newref{lem}{name = \RSlemtxt}}
  {}

\makeatother

\begin{document}
\title{Two-colour high-purity Einstein-Podolsky-Rosen photonic state}

\author{Tulio Brito Brasil}
\altaffiliation{These authors contributed equally to this work}
\affiliation{Niels Bohr Institute, University of Copenhagen, Copenhagen, Denmark}
\author{Valeriy Novikov}
\altaffiliation{These authors contributed equally to this work}
\affiliation{Niels Bohr Institute, University of Copenhagen, Copenhagen, Denmark}

\author{Hugo Kerdoncuff}
\affiliation{Danish Fundamental Metrology, Hørsholm, Denmark.}
\author{Mikael Lassen}
\affiliation{Danish Fundamental Metrology, Hørsholm, Denmark.}
\author{Eugene Polzik}
\affiliation{Niels Bohr Institute, University of Copenhagen, Copenhagen, Denmark}

\maketitle
\textbf{Entanglement is the backbone of quantum information science and its applications \cite{Nielsen2010}. Entangled states of light are necessary for distributed quantum protocols, quantum sensing \cite{Giovannetti2011} and quantum internet \cite{Kimble2008}. A distributed quantum network requires entanglement between light modes of different colour optimized for interaction with the nodes as well as for communication between them. 
Here we demonstrate high-purity Einstein-Podolsky-Rosen (EPR) entangled state between light modes with the wavelengths separated by more than $200$ nm. The modes display $-7.7\pm0.5$ dB of two-mode entanglement and an overall state purity of $0.63\pm0.16$. Entanglement is observed over five octaves of sideband frequencies from rf down to audio-band. In the context of two-colour entanglement, the demonstrated combination of high state purity, strong entanglement, and extended frequency range paves the way to new matter-light quantum protocols, such as  teleportation between disparate quantum systems \cite{Pirandola2015}, quantum sensing \cite{Giovannetti2011,Lawrie2019,Degen2017} and quantum-enhanced gravitational wave interferometry \cite{Khalili2018,Zeuthen2019}. The scheme demonstrated here can be readily applied towards entanglement between telecom wavelengths and atomic quantum
memories \cite{Duan2001,Hammerer2010}.}\par
\begin{figure*}[ht]
\includegraphics[width=1\linewidth]{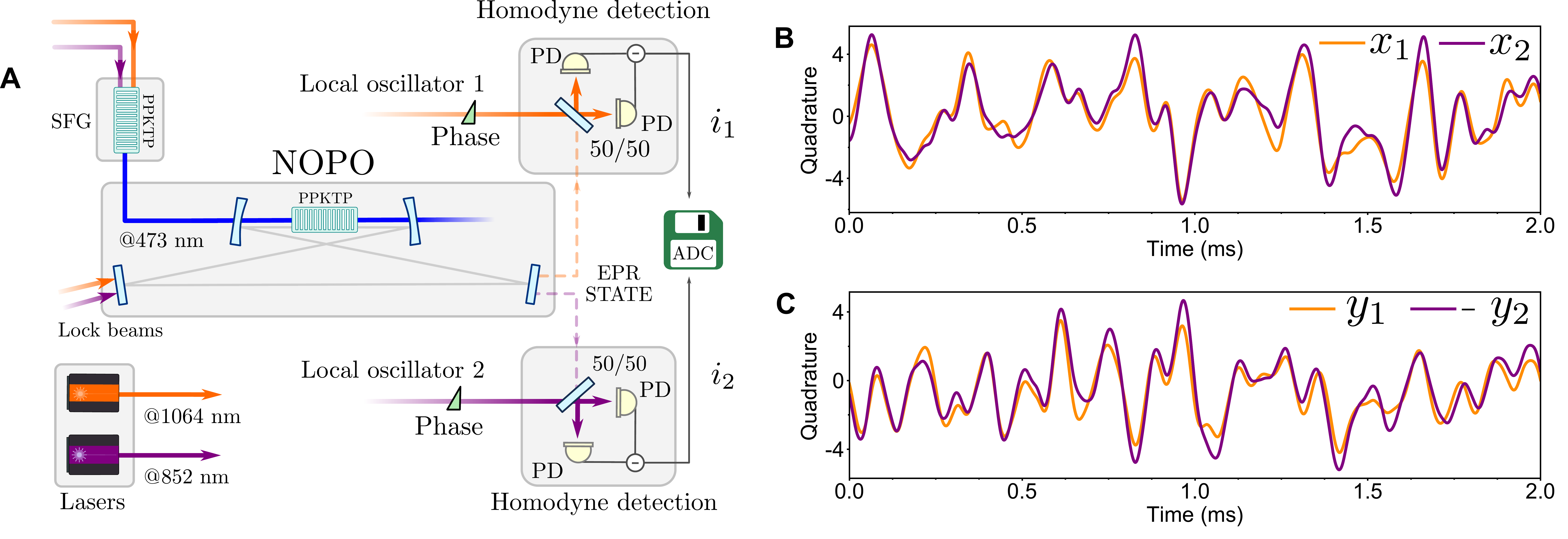}
\caption{(\textbf{A}) Scheme of the experimental setup. The 852 nm and 1064 nm lasers produce the local oscillators and the blue light used to pump the NOPO through the sum-frequency generation. The entangled modes at the two colours emerging from the NOPO are separated with a dichroic mirror, mixed with the LOs, and measured by the homodyne detectors. The photocurrents are recorded by the analog-to-digital converter (ADC) to obtain information on the joint system operators. (\textbf{B}-\textbf{C}) The experimental realizations of the photocurrents $i_1$ and $i_2$ showing strong non-classical correlations for $\{x_1,x_2\}$ and $\{y_1,-y_2\}$. Here the signals were demodulated at 200 kHz and integrated by a 10 kHz low-pass filter. The quadrature values are in vacuum state units.}
\label{fig:setup}
\end{figure*}
The nonlinear optics toolbox is the primary resource for generating photonic entangled states in continuous variables (CV). The $\chi^{(2)}$ nonlinearity is usually explored to produce single-mode squeezed states, where the noise in one quadrature is suppressed beyond the vacuum field fluctuations. The interference of single-mode squeezed states have been used to demonstrate strong two-mode entanglement \cite{Eberle2013,Steinlechner2013}, but this method limits the EPR states to be monochromatic.\par
As alternatives, second-harmonic generation (SHG) \cite{Grosse2008} and the process of non-degenerate parametric down-conversion \cite{Ou1992,Schori2002} allow for the generation of correlations between modes at different frequencies. The non-degenerate parametric process produces nonclassical correlation via annihilation of a photon with the frequency $\omega_0$ (pump) generating twin photons pairs with frequencies $\omega_1$ (signal) and $\omega_2$ (idler); satisfying $\omega_0 = \omega_1 + \omega_2$ and having the squeezed state production as the case where $\omega_1=\omega_2$. \par
Previous approaches to generation of multi-colour quantum correlations with frequency non-degenerate optical parametric oscillators (NOPO) differ significantly from the degenerate case due to the lack of frequency-matched local oscillators (LOs).  Operation above the oscillation threshold has been used to overcome this problem, but it increases the complexity of the system making it more susceptible to noise contamination and resulting in modest levels of entanglement \cite{Villar2005,Guo2012,Wang2018a}. Noteworthy, to the best of our knowledge, low-frequency (< 500 kHz) two-colour entanglement relevant for sensing applications has never been demonstrated.\par
Here we demonstrate a high-quality, tunable and versatile two-colour  entanglement source enabled by a novel experimental scheme. Two coherent laser sources are upconverted via the sum-frequency generation (SFG) $\omega_1 + \omega_2 = \omega_{\text{SFG}}$, and the output is  used as a pump beam for the NOPO below threshold, see Figure \ref{fig:setup}. The entangled output modes of the NOPO, $\Omega_{1,i}$ and $\Omega_{2,i}$ centered around two different colours of the lasers, $\omega_1$, $\omega_2$, respectively, are separated and superimposed with the coherent states at $\omega_1$ and $\omega_2$ enabling independent detection and entanglement verification. Optoelectronic control of the double-resonance NOPO and wide tunability of the relative phases of the four quadrature operators allow for generation of robust and strong two-colour EPR entanglement. The particular choice of the wavelengths of the entangled modes at 852 nm and 1064 nm has been motivated by the envisioned application for entanglement-enhanced gravitational wave interferometry \cite{Khalili2018,Zeuthen2019}.\par
To demonstrate the quantum correlations, we apply the EPR-paradox framework \cite{Einstein1935,Reid2009} of Reid's EPR criterion \cite{Reid1989}. In this context, we reproduce the original EPR-paradox situation if by measurements on one the subsystems one can infer the expected values of variables in the other subsystem in such a way as to obtain an apparent violation of the Heisenberg uncertainty principle. Consider noncommuting variables associated with the signal (1) and idler (2) field quadratures, $[x_j,y_j]=2i$, $j\,\in\, \{1,2\}$.  We take the violation of the inequality defined in \cite{Reid1989,Cavalcanti2009} as the measure of the EPR entanglement as follows
\begin{equation}
\label{eq1}
    \mathcal{E}=\Delta_{1|2}(x_1;x_2)\Delta_{1|2}(y_1;y_2) < 1,
\end{equation}
where the conditional uncertainty is defined as $\Delta_{1|2}(O_1;O_2)=\mathrm{min}_w\Delta(O_1-w_O O_2)$, with the parameter $w_O \in \mathbb{R}$. This was later recognized as an EPR-steering criterion and is sufficient and necessary for Gaussian states \cite{Cavalcanti2009}, ruling out the local state description of the system (1) or (2). Moreover, $\mathcal{E}^2$ can be used as a quantifier for the degree of EPR entanglement in a system \cite{Reid2009}.\par
Another necessary and sufficient entanglement criterion for Gaussian states is based on the inequality \cite{Simon2000,Duan2000}
\begin{align}\label{eq:EPR}
V^{\text{EPR}}=\text{Var}\left[\frac{x_1- x_2}{\sqrt{2}}\right] + \text{Var}\left[\frac{y_1 + y_2}{\sqrt{2}}\right] & < 2,
\end{align}
where the sum of variances $V^{\text{EPR}}$ allows finding the entanglement of formation (EoF) \cite{Wolf2004} which defines a number of e-bits that can be distilled from the state. \par
The theory comprising the dynamics of the EPR variables from a NOPO can be found in \cite{Drummond90,Schori2002}. In a nutshell, the generalized quadrature operator  $x_1(\theta) \equiv e^{i\theta}a_1^\dagger + e^{-i\theta}a_1$ is correlated with $x_2(-\theta)$, while $y_1(\theta)\equiv x_1(\theta+\pi/2)$ is anticorrelated with $y_2(-\theta)$, here the pump phase is taken as a reference. The variances of the two-mode operators $X^\pm(\theta) = [x_1(\theta) \pm x_2(-\theta)]/\sqrt{2}$, $Y^\pm(\theta)\equiv X^\pm(\theta+\pi/2)$  in case of symmetric losses are given by the well-known expressions \cite{Drummond90}
\begin{align}
\label{eq10}
V_{X^{\pm}} &= 1 \pm \eta^\mathrm{tot}\frac
{4\sqrt{\sigma}}{\tilde{\Omega}^{2}+(1\mp\sqrt{\sigma})^{2}},\\
V_{Y^{\pm}} &= 1 \mp \eta^\mathrm{tot}\frac
{4\sqrt{\sigma}}{\tilde{\Omega}^{2}+(1\pm\sqrt{\sigma})^{2}},
\end{align}
where $\sigma = P/P_{\mathrm{th}}$ is the pump power ($P$) normalized by the threshold power ($P_{\mathrm{th}}$), $\tilde{\Omega}=\Omega/\delta\nu$ is the measured noise sideband frequency ($\Omega$) normalized  by the cavity bandwidth ($\delta\nu$), and $\eta^{\mathrm{tot}}$ is the total efficiency \cite{Drummond90,Schori2002}. Thus the sum and the difference of the quadratures behave as two independent single-mode squeezed subspaces.\par
Several factors may affect the observation of optimum entanglement. Squeezing in $X^-$ and $Y^+$ is the signature of EPR correlations, but asymmetric losses may require optimization of the quadrature combination to achieve the best value of cross-correlations \cite{Schori2002}. Another limitation is due to the angular jitter of the noise ellipse, leading to a  projection of anti-squeezing onto the squeezed quadrature \cite{Oelker2016}. The effect of the phase noise of an arbitrary quadrature operator $Q(\theta)$ can be modeled by 
\begin{equation}
\label{eq:ph_noise}
V_{Q}(\delta\theta_n) = \cos^2 (\delta\theta_n) V_{Q(\theta)} +\sin^2 (\delta\theta_n) V_{Q(\theta+\pi/2)},\\
\end{equation}
where $\delta\theta_n$ is the RMS phase noise. Therefore, for a finite $\delta\theta_n$ the highest level of noise reduction is achieved at a specific pump power. Because phase noise is always present, below we use  $V_{Q}$ and $V_{Q}(\delta\theta_n)$ interchangeably.\par

The layout of the experimental setup is presented in Fig. \ref{fig:setup}a. We measure the field quadratures from the NOPO output to observe correlations and determine the EPR entanglement between the signal and idler beams. Each of the two entangled modes is directed to a balanced homodyne detector to be mixed with the corresponding local oscillator --- control of the relative phase between the local oscillator and the entangled mode selects which quadrature is projected into the photocurrents $i_1 \varpropto x_1(\theta_1)$ and $i_2\varpropto x_2(\theta_2)$. Figs. \ref{fig:setup} (b-c) show the experimental realizations of the photocurrents presenting strong non-classical correlations between the quadrature measurements by the two detectors.\par

\begin{figure*}[ht]
\begin{center}
\includegraphics[width=\linewidth]{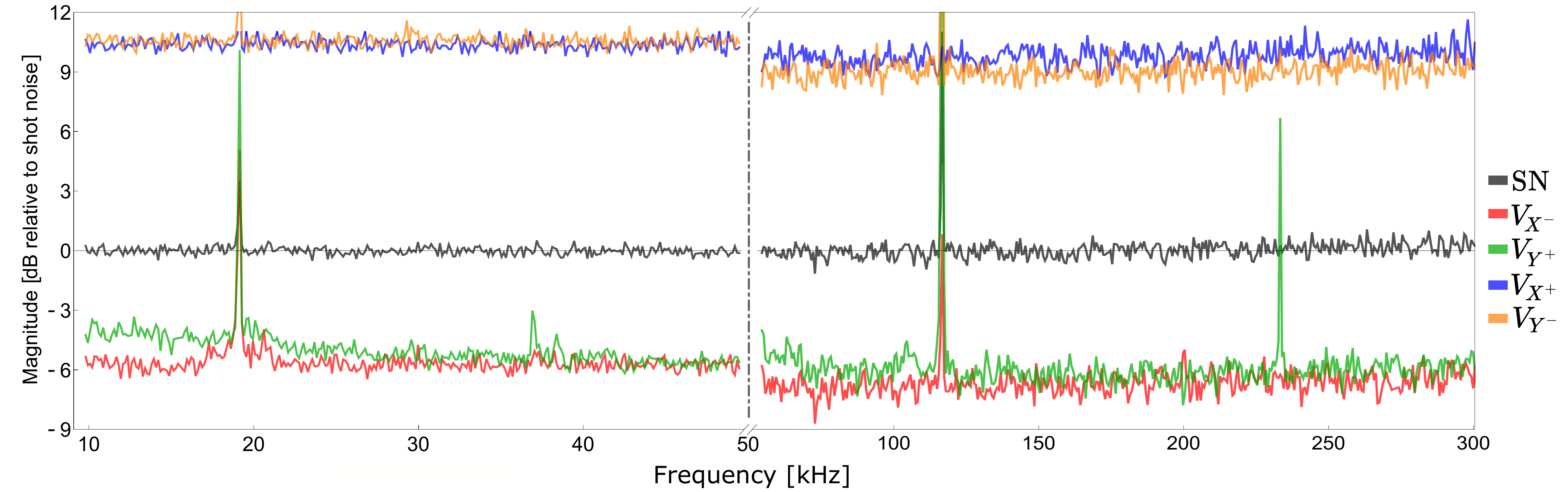} 
\caption{Spectra of the EPR quadratures normalized to shot-noise level (SN) for the frequency range 10 to 300 kHz. The left traces show the quantum noise suppression optimized for low spectral frequencies (10-50 kHz), while the right part corresponds to the best entanglement level achieved in 50-300 kHz spectral range (see text below). The narrow peaks come from the phase noise of the lasers. The data are corrected for electronic noise which is $18.5$ dB below the shot-noise level.}
\label{fig:measurement_results_combined}
\end{center}
\end{figure*}
The relative phase $\theta_j = \phi_j^{\text{OPO}}-\phi_j^{\text{LO}}$ is recovered from the interference between the local oscillator and a weak back reflection of the cavity lock beam (Fig. \ref{fig:setup}a) from the intracavity nonlinear crystal. We use this interference information to control the phase in each LO path with a PZT and to select which quadrature to measure (see Supplementary Information). The observables $X^{\pm}$ ($Y^{\pm}$) are recorded by initially setting of $\theta_1 = \theta_2 = 0 $ $(\pi/2)$ and the subsequent fine adjustment of one of the phases to maximize the measured correlations.\par
We found the optimal pumping power to be $\sigma\approx$ 0.25, corresponding to operation well below the threshold. For these pumping conditions, we measured $V_{X^-}=-7.1\pm0.5 $ dB; $V_{Y^+}=-6.2\pm0.5 $ dB for the frequency range 50 to 300 kHz (see Fig. \ref{fig:measurement_results_combined}). 
Increasing the pump power does not improve the entanglement level due to the enhanced phase noise influence. Further down in the audio frequency band, the entanglement is extremely vulnerable to phase noise. The combination of passive stability and active optoelectronic control allows us to achieve entanglement down to below $10$ kHz (Fig. \ref{fig:measurement_results_combined}). However, even the optimized low-frequency noise limited the EPR correlations to $V_{X^-}= -5.7\pm0.6$ dB;  $V_{Y^+}=-5.2\pm0.6$ dB. \par

We obtain the quadratures of the optical field $V_Q^\text{o}$ characterizing its entanglement by correcting the measured variances $V_Q$ by the non-unity quantum efficiency of the detectors  $\eta^\text{det}$ as
\begin{equation}
    V_Q =  \eta^{\text{det}}(V_Q^{\text{o}}-1)+1, 
\end{equation}
Using the average detector efficiency \cite{Schori2002} (see Methods) $\eta^\mathrm{det}=\sqrt{\eta_1^\mathrm{det}\eta_2^\mathrm{det}}=0.945$ , we obtain the following variances of the light modes for the data presented in Fig. \ref{fig:measurement_results_combined}: $V^ {\text{o}}_{X^-}= -8.3\pm0.6$ dB; $V^{\text{o}}_{Y^+} = -7.1\pm0.5$ dB and $V^{\text{o}}_{X^+}=10.0\pm0.5 $ dB; $V^{\text{o}}_{Y^-} = 9.3\pm0.6$ dB. From that we obtain the entanglement measures for the optical fields: $\mathcal{E}^{2}_{\text{c}}=0.029\pm 0.007 \ll 1$ and $V^{\text{EPR,c}}= 0.34\pm 0.04 \ll 2$ ($-7.7\pm0.5$ dB below the entanglement boundary). To the best of our knowledge, this is the highest level of EPR entanglement achieved between beams at disparate wavelengths. \par
To fully characterize our EPR state, we focus on its purity, which is directly related to the twin-photon nature of the parametric down-conversion process. For a Gaussian state with the covariance matrix $\mathbb{V}$ the state purity is given by $\mu=1/\sqrt{\text{Det} \mathbb{V}}=1/\sqrt{V^{\text{o}}_{X^-}V^{\text{o}}_{Y^-}V^{\text{o}}_{X^+}V^{\text{o}}_{Y^+}}$\cite{Adesso2014} which yields $\mu = 0.63 \pm0.16$.  Our results compare favorably with the highest  to date two-colour entanglement with $V^{\text{EPR},c}= 0.38 $ and with purity $0.11$ (corrected by the reported $\eta^{\text{det}}$) observed in the MHz range \cite{Wang2018a}. High purity is especially relevant for quantum enhancement of interferometry where both quadratures can contain useful information \cite{Yu2020,Steinlechner2013a}.\par

We have presented experimental realization of the EPR entangled state of light between modes of different colours with unprecedented degree of entanglement and purity and have expanded the entanglement into the acoustic frequency range.\par
The present limitations to the observed entanglement are primarily due to the imperfect phase-locking affected by spurious non-linear mixing. Improvement of phase control would make possible to observe even higher levels of entanglement, preserving the state purity. Changing the phase control scheme to a coherent phase-lock \cite{Vahlbruch2006} and locking the cavity by frequency-shifted modes would eliminate classical noise injection, making it possible to observe two-colour entanglement down to the Hz domain.\par
The present choice of wavelengths of $1064$ nm and $852$ nm is motivated by the current laser choice of LIGO and is geared towards the broadband quantum noise reduction in gravitational wave detectors using an ensemble of Caesium atoms \cite{Khalili2018,Zeuthen2019}. Our approach is readily applicable to entanglement generation between other colours, for example, between $852$ nm compatible with the atomic quantum memory and a telecom wavelength.

\bibliographystyle{naturemag}
\bibliography{references.bib}

\begin{thebibliography}{10}
\expandafter\ifx\csname url\endcsname\relax
  \def\url#1{\texttt{#1}}\fi
\expandafter\ifx\csname urlprefix\endcsname\relax\def\urlprefix{URL }\fi
\providecommand{\bibinfo}[2]{#2}
\providecommand{\eprint}[2][]{\url{#2}}

\bibitem{Nielsen2010}
\bibinfo{author}{Nielsen, M.~A.} \& \bibinfo{author}{Chuang, I.~L.}
\newblock \emph{\bibinfo{title}{Quantum Computation and Quantum Information:
  10th Anniversary Edition}} (\bibinfo{publisher}{Cambridge University Press},
  \bibinfo{year}{2010}).
\newblock \urlprefix\url{http://10.1017/CBO9780511976667}.

\bibitem{Giovannetti2011}
\bibinfo{author}{Giovannetti, V.}, \bibinfo{author}{Lloyd, S.} \&
  \bibinfo{author}{Maccone, L.}
\newblock \bibinfo{title}{Advances in quantum metrology}.
\newblock \emph{\bibinfo{journal}{Nature Photonics}}
  \textbf{\bibinfo{volume}{5}}, \bibinfo{pages}{222--229}
  (\bibinfo{year}{2011}).
\newblock \urlprefix\url{https://doi.org/10.1038/nphoton.2011.35}.

\bibitem{Kimble2008}
\bibinfo{author}{Kimble, H.~J.}
\newblock \bibinfo{title}{The quantum internet}.
\newblock \emph{\bibinfo{journal}{Nature}} \textbf{\bibinfo{volume}{453}},
  \bibinfo{pages}{1023--1030} (\bibinfo{year}{2008}).
\newblock \urlprefix\url{https://doi.org/10.1038/nature07127}.

\bibitem{Pirandola2015}
\bibinfo{author}{Pirandola, S.}, \bibinfo{author}{Eisert, J.},
  \bibinfo{author}{Weedbrook, C.}, \bibinfo{author}{Furusawa, A.} \&
  \bibinfo{author}{Braunstein, S.~L.}
\newblock \bibinfo{title}{{Advances in quantum teleportation}}.
\newblock \emph{\bibinfo{journal}{Nature Photonics}}
  \textbf{\bibinfo{volume}{9}}, \bibinfo{pages}{641--652}.
\newblock \urlprefix\url{https://www.nature.com/articles/nphoton.2015.154}.

\bibitem{Lawrie2019}
\bibinfo{author}{Lawrie, B.~J.}, \bibinfo{author}{Lett, P.~D.},
  \bibinfo{author}{Marino, A.~M.} \& \bibinfo{author}{Pooser, R.~C.}
\newblock \bibinfo{title}{Quantum sensing with squeezed light}.
\newblock \emph{\bibinfo{journal}{ACS Photonics}} \textbf{\bibinfo{volume}{6}},
  \bibinfo{pages}{1307--1318} (\bibinfo{year}{2019}).
\newblock \urlprefix\url{https://doi.org/10.1021/acsphotonics.9b00250}.

\bibitem{Degen2017}
\bibinfo{author}{Degen, C.~L.}, \bibinfo{author}{Reinhard, F.} \&
  \bibinfo{author}{Cappellaro, P.}
\newblock \bibinfo{title}{Quantum sensing}.
\newblock \emph{\bibinfo{journal}{Rev. Mod. Phys.}}
  \textbf{\bibinfo{volume}{89}}, \bibinfo{pages}{035002}
  (\bibinfo{year}{2017}).
\newblock
  \urlprefix\url{https://link.aps.org/doi/10.1103/RevModPhys.89.035002}.

\bibitem{Khalili2018}
\bibinfo{author}{Khalili, F.~Y.} \& \bibinfo{author}{Polzik, E.~S.}
\newblock \bibinfo{title}{Overcoming the standard quantum limit in
  gravitational wave detectors using spin systems with a negative effective
  mass}.
\newblock \emph{\bibinfo{journal}{Phys. Rev. Lett.}}
  \textbf{\bibinfo{volume}{121}}, \bibinfo{pages}{031101}
  (\bibinfo{year}{2018}).
\newblock
  \urlprefix\url{https://link.aps.org/doi/10.1103/PhysRevLett.121.031101}.

\bibitem{Zeuthen2019}
\bibinfo{author}{Zeuthen, E.}, \bibinfo{author}{Polzik, E.~S.} \&
  \bibinfo{author}{Khalili, F.~Y.}
\newblock \bibinfo{title}{Gravitational wave detection beyond the standard
  quantum limit using a negative-mass spin system and virtual rigidity}.
\newblock \emph{\bibinfo{journal}{Phys. Rev. D}}
  \textbf{\bibinfo{volume}{100}}, \bibinfo{pages}{062004}
  (\bibinfo{year}{2019}).
\newblock \urlprefix\url{https://link.aps.org/doi/10.1103/PhysRevD.100.062004}.

\bibitem{Duan2001}
\bibinfo{author}{Duan, L.~M.}, \bibinfo{author}{Lukin, M.~D.},
  \bibinfo{author}{Cirac, J.~I.} \& \bibinfo{author}{Zoller, P.}
\newblock \bibinfo{title}{{Long-distance quantum communication with atomic
  ensembles and linear optics}}.
\newblock \emph{\bibinfo{journal}{Nature}} \textbf{\bibinfo{volume}{414}},
  \bibinfo{pages}{413--418}.
\newblock \urlprefix\url{https://www.nature.com/articles/35106500}.

\bibitem{Hammerer2010}
\bibinfo{author}{Hammerer, K.}, \bibinfo{author}{S\o{}rensen, A.~S.} \&
  \bibinfo{author}{Polzik, E.~S.}
\newblock \bibinfo{title}{Quantum interface between light and atomic
  ensembles}.
\newblock \emph{\bibinfo{journal}{Rev. Mod. Phys.}}
  \textbf{\bibinfo{volume}{82}}, \bibinfo{pages}{1041--1093}
  (\bibinfo{year}{2010}).
\newblock \urlprefix\url{https://link.aps.org/doi/10.1103/RevModPhys.82.1041}.

\bibitem{Eberle2013}
\bibinfo{author}{Eberle, T.}, \bibinfo{author}{H\"{a}ndchen, V.} \&
  \bibinfo{author}{Schnabel, R.}
\newblock \bibinfo{title}{Stable control of 10 \text{dB} two-mode squeezed
  vacuum states of light}.
\newblock \emph{\bibinfo{journal}{Opt. Express}} \textbf{\bibinfo{volume}{21}},
  \bibinfo{pages}{11546--11553} (\bibinfo{year}{2013}).
\newblock
  \urlprefix\url{http://www.opticsexpress.org/abstract.cfm?URI=oe-21-9-11546}.

\bibitem{Steinlechner2013}
\bibinfo{author}{Steinlechner, S.}, \bibinfo{author}{Bauchrowitz, J.},
  \bibinfo{author}{Eberle, T.} \& \bibinfo{author}{Schnabel, R.}
\newblock \bibinfo{title}{Strong \text{E}instein-\text{P}odolsky-\text{R}osen
  steering with unconditional entangled states}.
\newblock \emph{\bibinfo{journal}{Phys. Rev. A}} \textbf{\bibinfo{volume}{87}},
  \bibinfo{pages}{022104} (\bibinfo{year}{2013}).
\newblock \urlprefix\url{https://link.aps.org/doi/10.1103/PhysRevA.87.022104}.

\bibitem{Grosse2008}
\bibinfo{author}{Grosse, N.~B.} \emph{et~al.}
\newblock \bibinfo{title}{Observation of entanglement between two light beams
  spanning an octave in optical frequency}.
\newblock \emph{\bibinfo{journal}{Phys. Rev. Lett.}}
  \textbf{\bibinfo{volume}{100}}, \bibinfo{pages}{243601}
  (\bibinfo{year}{2008}).
\newblock
  \urlprefix\url{https://link.aps.org/doi/10.1103/PhysRevLett.100.243601}.

\bibitem{Ou1992}
\bibinfo{author}{Ou, Z.~Y.}, \bibinfo{author}{Pereira, S.~F.},
  \bibinfo{author}{Kimble, H.~J.} \& \bibinfo{author}{Peng, K.~C.}
\newblock \bibinfo{title}{Realization of the
  \text{E}instein-\text{P}odolsky-\text{R}osen paradox for continuous
  variables}.
\newblock \emph{\bibinfo{journal}{Phys. Rev. Lett.}}
  \textbf{\bibinfo{volume}{68}}, \bibinfo{pages}{3663--3666}
  (\bibinfo{year}{1992}).
\newblock \urlprefix\url{https://link.aps.org/doi/10.1103/PhysRevLett.68.3663}.

\bibitem{Schori2002}
\bibinfo{author}{Schori, C.}, \bibinfo{author}{S\o{}rensen, J.~L.} \&
  \bibinfo{author}{Polzik, E.~S.}
\newblock \bibinfo{title}{Narrow-band frequency tunable light source of
  continuous quadrature entanglement}.
\newblock \emph{\bibinfo{journal}{Phys. Rev. A}} \textbf{\bibinfo{volume}{66}},
  \bibinfo{pages}{033802} (\bibinfo{year}{2002}).
\newblock \urlprefix\url{https://link.aps.org/doi/10.1103/PhysRevA.66.033802}.

\bibitem{Villar2005}
\bibinfo{author}{Villar, A.~S.}, \bibinfo{author}{Cruz, L.~S.},
  \bibinfo{author}{Cassemiro, K.~N.}, \bibinfo{author}{Martinelli, M.} \&
  \bibinfo{author}{Nussenzveig, P.}
\newblock \bibinfo{title}{Generation of bright two-color continuous variable
  entanglement}.
\newblock \emph{\bibinfo{journal}{Phys. Rev. Lett.}}
  \textbf{\bibinfo{volume}{95}}, \bibinfo{pages}{243603}
  (\bibinfo{year}{2005}).
\newblock
  \urlprefix\url{https://link.aps.org/doi/10.1103/PhysRevLett.95.243603}.

\bibitem{Guo2012}
\bibinfo{author}{Guo, X.}, \bibinfo{author}{Zhao, J.} \& \bibinfo{author}{Li,
  Y.}
\newblock \bibinfo{title}{Robust generation of bright two-color entangled
  optical beams from a phase-insensitive optical parametric amplifier}.
\newblock \emph{\bibinfo{journal}{Applied Physics Letters}}
  \textbf{\bibinfo{volume}{100}}, \bibinfo{pages}{091112}
  (\bibinfo{year}{2012}).
\newblock \urlprefix\url{https://doi.org/10.1063/1.3690876}.
\newblock \eprint{https://doi.org/10.1063/1.3690876}.

\bibitem{Wang2018a}
\bibinfo{author}{Wang, N.}, \bibinfo{author}{Du, S.} \& \bibinfo{author}{Li,
  Y.}
\newblock \bibinfo{title}{Compact 6 \text{dB} two-color continuous variable
  entangled source based on a single ring optical resonator}.
\newblock \emph{\bibinfo{journal}{Applied Sciences}}
  \textbf{\bibinfo{volume}{8}}, \bibinfo{pages}{330} (\bibinfo{year}{2018}).
\newblock \urlprefix\url{https://www.mdpi.com/2076-3417/8/3/330}.

\bibitem{Einstein1935}
\bibinfo{author}{Einstein, A.}, \bibinfo{author}{Podolsky, B.} \&
  \bibinfo{author}{Rosen, N.}
\newblock \bibinfo{title}{Can quantum-mechanical description of physical
  reality be considered complete?}
\newblock \emph{\bibinfo{journal}{Phys. Rev.}} \textbf{\bibinfo{volume}{47}},
  \bibinfo{pages}{777--780} (\bibinfo{year}{1935}).
\newblock \urlprefix\url{https://link.aps.org/doi/10.1103/PhysRev.47.777}.

\bibitem{Reid2009}
\bibinfo{author}{Reid, M.~D.} \emph{et~al.}
\newblock \bibinfo{title}{Colloquium: The einstein-podolsky-rosen paradox: From
  concepts to applications}.
\newblock \emph{\bibinfo{journal}{Rev. Mod. Phys.}}
  \textbf{\bibinfo{volume}{81}}, \bibinfo{pages}{1727--1751}
  (\bibinfo{year}{2009}).
\newblock \urlprefix\url{https://link.aps.org/doi/10.1103/RevModPhys.81.1727}.

\bibitem{Reid1989}
\bibinfo{author}{Reid, M.~D.}
\newblock \bibinfo{title}{Demonstration of the
  \text{E}instein-\text{P}odolsky-\text{R}osen paradox using nondegenerate
  parametric amplification}.
\newblock \emph{\bibinfo{journal}{Phys. Rev. A}} \textbf{\bibinfo{volume}{40}},
  \bibinfo{pages}{913--923} (\bibinfo{year}{1989}).
\newblock \urlprefix\url{https://link.aps.org/doi/10.1103/PhysRevA.40.913}.

\bibitem{Cavalcanti2009}
\bibinfo{author}{Cavalcanti, E.~G.}, \bibinfo{author}{Jones, S.~J.},
  \bibinfo{author}{Wiseman, H.~M.} \& \bibinfo{author}{Reid, M.~D.}
\newblock \bibinfo{title}{Experimental criteria for steering and the
  \text{E}instein-\text{P}odolsky-\text{R}osen paradox}.
\newblock \emph{\bibinfo{journal}{Phys. Rev. A}} \textbf{\bibinfo{volume}{80}},
  \bibinfo{pages}{032112} (\bibinfo{year}{2009}).
\newblock \urlprefix\url{https://link.aps.org/doi/10.1103/PhysRevA.80.032112}.

\bibitem{Simon2000}
\bibinfo{author}{Simon, R.}
\newblock \bibinfo{title}{\text{P}eres-\text{H}orodecki separability criterion
  for continuous variable systems}.
\newblock \emph{\bibinfo{journal}{Phys. Rev. Lett.}}
  \textbf{\bibinfo{volume}{84}}, \bibinfo{pages}{2726--2729}
  (\bibinfo{year}{2000}).
\newblock \urlprefix\url{https://link.aps.org/doi/10.1103/PhysRevLett.84.2726}.

\bibitem{Duan2000}
\bibinfo{author}{Duan, L.-M.}, \bibinfo{author}{Giedke, G.},
  \bibinfo{author}{Cirac, J.~I.} \& \bibinfo{author}{Zoller, P.}
\newblock \bibinfo{title}{Inseparability criterion for continuous variable
  systems}.
\newblock \emph{\bibinfo{journal}{Phys. Rev. Lett.}}
  \textbf{\bibinfo{volume}{84}}, \bibinfo{pages}{2722--2725}
  (\bibinfo{year}{2000}).
\newblock \urlprefix\url{https://link.aps.org/doi/10.1103/PhysRevLett.84.2722}.

\bibitem{Wolf2004}
\bibinfo{author}{Wolf, M.~M.}, \bibinfo{author}{Giedke, G.},
  \bibinfo{author}{Kr\"uger, O.}, \bibinfo{author}{Werner, R.~F.} \&
  \bibinfo{author}{Cirac, J.~I.}
\newblock \bibinfo{title}{Gaussian entanglement of formation}.
\newblock \emph{\bibinfo{journal}{Phys. Rev. A}} \textbf{\bibinfo{volume}{69}},
  \bibinfo{pages}{052320} (\bibinfo{year}{2004}).
\newblock \urlprefix\url{https://link.aps.org/doi/10.1103/PhysRevA.69.052320}.

\bibitem{Drummond90}
\bibinfo{author}{Drummond, P.~D.} \& \bibinfo{author}{Reid, M.~D.}
\newblock \bibinfo{title}{Correlations in nondegenerate parametric oscillation.
  \text{II}. below threshold results}.
\newblock \emph{\bibinfo{journal}{Phys. Rev. A}} \textbf{\bibinfo{volume}{41}},
  \bibinfo{pages}{3930--3949} (\bibinfo{year}{1990}).
\newblock \urlprefix\url{https://link.aps.org/doi/10.1103/PhysRevA.41.3930}.

\bibitem{Oelker2016}
\bibinfo{author}{Oelker, E.} \emph{et~al.}
\newblock \bibinfo{title}{Ultra-low phase noise squeezed vacuum source for
  gravitational wave detectors}.
\newblock \emph{\bibinfo{journal}{Optica}} \textbf{\bibinfo{volume}{3}},
  \bibinfo{pages}{682--685} (\bibinfo{year}{2016}).
\newblock
  \urlprefix\url{http://www.osapublishing.org/optica/abstract.cfm?URI=optica-3-7-682}.

\bibitem{Adesso2014}
\bibinfo{author}{Adesso, G.}, \bibinfo{author}{Ragy, S.} \&
  \bibinfo{author}{Lee, A.~R.}
\newblock \bibinfo{title}{Continuous variable quantum information: Gaussian
  states and beyond}.
\newblock \emph{\bibinfo{journal}{Open Systems \& Information Dynamics}}
  \textbf{\bibinfo{volume}{21}}, \bibinfo{pages}{1440001}
  (\bibinfo{year}{2014}).
\newblock \urlprefix\url{https://doi.org/10.1142/S1230161214400010}.

\bibitem{Yu2020}
\bibinfo{author}{Yu, H.} \emph{et~al.}
\newblock \bibinfo{title}{Quantum correlations between light and the
  kilogram-mass mirrors of ligo}.
\newblock \emph{\bibinfo{journal}{Nature}} \textbf{\bibinfo{volume}{583}},
  \bibinfo{pages}{43--47} (\bibinfo{year}{2020}).
\newblock \urlprefix\url{https://doi.org/10.1038/s41586-020-2420-8}.

\bibitem{Steinlechner2013a}
\bibinfo{author}{Steinlechner, S.} \emph{et~al.}
\newblock \bibinfo{title}{Quantum-dense metrology}.
\newblock \emph{\bibinfo{journal}{Nature Photonics}}
  \textbf{\bibinfo{volume}{7}}, \bibinfo{pages}{626--630}
  (\bibinfo{year}{2013}).
\newblock \urlprefix\url{https://doi.org/10.1038/nphoton.2013.150}.

\bibitem{Vahlbruch2006}
\bibinfo{author}{Vahlbruch, H.} \emph{et~al.}
\newblock \bibinfo{title}{Coherent control of vacuum squeezing in the
  gravitational-wave detection band}.
\newblock \emph{\bibinfo{journal}{Phys. Rev. Lett.}}
  \textbf{\bibinfo{volume}{97}}, \bibinfo{pages}{011101}
  (\bibinfo{year}{2006}).
\newblock
  \urlprefix\url{https://link.aps.org/doi/10.1103/PhysRevLett.97.011101}.

\bibitem{Mabuchi1994}
\bibinfo{author}{Mabuchi, H.}, \bibinfo{author}{Polzik, E.~S.} \&
  \bibinfo{author}{Kimble, H.~J.}
\newblock \bibinfo{title}{Blue-light-induced infrared absorption in
  \text{KNbO}$_3$}.
\newblock \emph{\bibinfo{journal}{J. Opt. Soc. Am. B}}
  \textbf{\bibinfo{volume}{11}}, \bibinfo{pages}{2023--2029}
  (\bibinfo{year}{1994}).
\newblock
  \urlprefix\url{http://josab.osa.org/abstract.cfm?URI=josab-11-10-2023}.

\bibitem{Neuhaus2017}
\bibinfo{author}{Neuhaus, L. e.~a.}
\newblock \bibinfo{title}{Pyrpl (python red pitaya lockbox) — an open-source
  software package for fpga-controlled quantum optics experiments}.
\newblock In \emph{\bibinfo{booktitle}{2017 Conference on Lasers and
  Electro-Optics Europe European Quantum Electronics Conference
  (CLEO/Europe-EQEC)}}, \bibinfo{pages}{1--1} (\bibinfo{year}{2017}).
\newblock \urlprefix\url{https://ieeexplore.ieee.org/document/8087380}.

\end{thebibliography}

\clearpage

\section*{Methods}
\subsection*{NOPO design}
The NOPO cavity is designed and tuned to be resonant for both signal (852 nm) and idler (1064 nm) beams while the pump beam (473 nm) is used in a single-pass regime. The cavity has a bow-tie configuration to reduce the negative influence of back-scattered light and to improve the escape efficiency \cite{Ou1992}. Quantum light emerges through the output coupling mirror with the transmission coefficient $T=12\%$ for both  852 nm and 1064 nm modes. Thus, the cavity bandwidth, free spectral range, and finesse are very similar for both wavelengths. The main NOPO parameters are given in Table \ref{tab:opo}.  To minimize astigmatism and contamination from high-order transverse modes, we fine-tune the cavity size and angles of incidence on the mirrors. The NOPO is built in a monolithic aluminum box for better mechanical stability. We use a type-0 periodically poled KTP (PPKTP) crystal (Raicol Crystals Ltd) as the nonlinear medium with an antireflection (AR) coating for 473 nm, 852 nm and, 1064 nm. The desired phase matching is achieved by setting the crystal temperature to $\approx$ 63 ${^\circ}$C and stabilizing it to $\pm$ 1 mK. The passive intracavity losses $\mathcal{L}_j$ are dominated by the PPKTP bulk losses, Table \ref{tab:opo}.\par
\begin{table}[ht]
\caption{NOPO main parameters.}
\label{tab:opo}
\begin{center}
\begin{tabular}{l l l l}
\hline
Parameter & & & Value    \\ \hline
Intracavity loss for 1064 nm  &  & &$0.15\pm 0.02\,$\% \\
Intracavity loss for 852 nm &  & &$0.21\pm0.02\,$\% \\
Cavity Length &  & & 390 mm\\
Mirror radius of curvature &  & & -38 mm\\
Free spectral range &  & & 769 MHz \\
Bandwidth &  & & 15 MHz\\
Finesse &  & & 52 \\
Threshold power (473 nm) & &  & $320\pm16$ mW\\
PPKTP dimensions &  &  &$1\times1\times10$ mm$^3$\\\hline
\end{tabular}
\end{center}
\end{table}
\subsection*{Estimated efficiencies}
The measured efficiencies in our system are shown in Table \ref{tab:eff}. The single beam escape efficiency is given by $\eta_j^{\text{esc}} = T/(T+\mathcal{L}_j)$, and is the most significant parameter to guarantee high-purity state generation. This leads to an overall escape efficiency $\eta^\text{esc} =\sqrt{\eta_1^ {\text{esc}}\eta_2^{\text{esc}}} = 98.5\pm 0.2\%$ \cite{Schori2002}. We also studied the effect of the blue-light-induced infrared absorption (BLIIRA) \cite{Mabuchi1994} on the overall escape efficiency and found it negligible under the normal pumping conditions. The combination of ultra-low intracavity losses and BLIIRA-free operation are unprecedented for a two-colour system giving an escape efficiency comparable to the state-of-art degenerate OPO \cite{Eberle2013}.\par
Table \ref{tab:eff} presents the propagation efficiency $\eta_j^{\text{pro}}$ from the NOPO output to the detectors, the homodyne efficiency $\eta^\mathrm{mm}$ of the  signal-LO mode-matching, and the photodiodes' quantum efficiency $\eta^\mathrm{det}_j$ (see Supplementary Information).

\begin{table}[ht]
\caption{Estimated efficiencies.}
\label{tab:eff}
\begin{center}
\begin{tabular}{ c c c c c c c c c }
\hline
$\lambda$ (nm) & &$\eta^\mathrm{esc}$ & &$\eta^\mathrm{pro}$ & &$\eta^\mathrm{mm}$ & &$\eta^\mathrm{det}$\\ \hline
\parbox[c]{1cm}{1064} & & $98.7\pm 0.1\%$ & & $99.1\pm 0.3\%$ & & $98.9\pm 1.1\%$ & &$93 \pm 2 \%$\\
\parbox[c]{1cm}{852} & &$98.3\pm 0.1\%$ & & $99.0\pm 0.3\%$ &  & $98.4\pm 1.5\%$ & &$96 \pm 2\%$ \\
\hline
\end{tabular}
\end{center}
\end{table}

\section*{Acknowledgments}
We gratefully acknowledge conversations with J. H. Müller and J. Appel.  This project has been funded by the InnoFond Denmark through the Eureka Turbo program, by the European Research Council Advanced grant QUANTUM-N, and by the Villum Foundation. T. B. B. was partially supported by CAPES and CNPQ.

\section*{Author Contributions}
E.\,S.\,P.\ conceived and led the project. 
T.B.B. and V.N. designed and built the experiment with the help of H.K. and M.L. T.B.B. and V.N. obtained the main experimental results.
The paper was written by E.\,S.\,P., T.B.B. and V.N., with contributions from H.K. and M.L. E.\,S.\,P.\ and M.L. supervised the research.

\section*{Author Information}
 The authors declare no competing financial 
interests. Correspondence and 
requests for materials should be addressed to T.B.B. (tulio.brasil@nbi.ku.dk ) or E.S.P. (polzik@nbi.ku.dk).

\section*{Data Availability Statement}
The data that support the findings of this study are available from the corresponding author upon reasonable request.


\appendix
\clearpage

\setcounter{page}{1}
\renewcommand{\thepage}{SI~\arabic{page}}

\setcounter{figure}{0}
\renewcommand{\thefigure}{SI\arabic{figure}}

\setcounter{table}{0}
\renewcommand{\thetable}{SI\arabic{table}}

\setcounter{equation}{0}
\renewcommand{\theequation}{SI~\thesection.\arabic{equation}}

\clearpage
\onecolumngrid
\section*{Supplementary Information}

\subsection*{Lasers and NOPO lock}
\begin{figure*}[ht]
\begin{center}
\includegraphics[width=\linewidth]{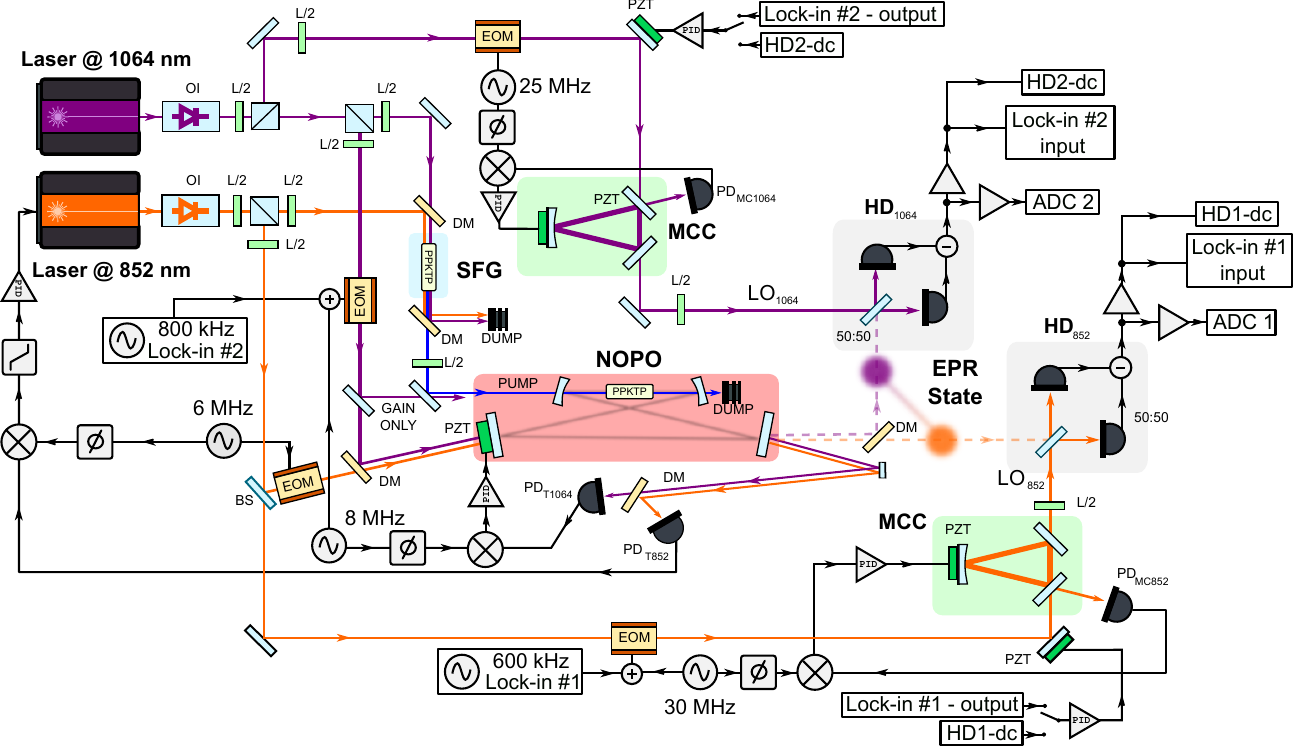} 
\caption{Detailed scheme of the two-colour entanglement experimental setup. The 852 nm (Ti:Sapphire) and 1064 nm (Mephisto + fiber amplifier) lasers produce the local oscillators and are up-converted to generate the pump field (473 nm). The twin beams from the NOPO output are separated by a dichroic mirror, mixed with the LOs, and sent to the homodyne detectors. The photodetectors PD$_{\text{T1064}}$ and PD$_{\text{T852}}$ are used to lock the NOPO cavity on double resonance, while PD$_{\text{MC1064}}$ and PD$_{\text{MC852}}$ are used to lock the mode-cleaner cavities. The homodyne photocurrents are split to be recorded by the analog-to-digital converter (ADC) and the phase lock scheme (DC and input of the lock-in amplifier). L/2 : half-wave plate. DM: dichroic mirror. BS: beam-splitter. MMC: Mode Cleaner Cavity. PID: Proportional-Integral-Differential control loop. OI: optical isolator. PD: photodetector. EOM: Electro-optical Modulator. PZT: Piezoelectric Transducer. }
\label{fig:setup_detailed}
\end{center}
\end{figure*}

A detailed scheme of the experimental implementation can be seen in \ref{fig:setup_detailed}. The laser at 1064 nm is a Nd:YAG (Innolight, Mephisto E500) with 500 mW of output power used to seed a fiber amplifier (NuFern, NUA-1064-50-10W-2-1), giving 10 W of output power. The laser at 852 nm is a Ti:Sapphire (Msquared,  PSX) with a maximum output power of 2.2 W. A crystal similar to the one used in the OPO is used to up-convert the initial lasers by single-pass SFG and producing the pump field at 473 nm. To keep the NOPO on double resonance, we inject weak beams,  counter-propagating relative to the output modes, through a high-reflectivity (HR) mirror and apply the Pound-Drever-Hall scheme using the transmitted light. The modulation frequencies used to generate the error signals are shown in \ref{fig:setup_detailed}.  The lock has two steps; first, we lock the cavity length relative to the 1064 nm laser --- for that, feedback is sent to a piezoelectric transducer (PZT) attached to one of the NOPO mirrors. Subsequently, we lock the 852 nm laser to the cavity by applying slow feedback to the PZT inside the Ti:Sapphire laser. In the end, the stability of the system is determined by frequency drifts of the 1064 nm laser and undesired mechanical and thermal fluctuations. All  PID feedback loops are controlled using FPGA boards (Redpitaya STEMLAB, 125-14) and the PyRPL package \cite{Neuhaus2017}.\par

When locking the homodyne angles, we switch between the two available error signals to feedback the local oscillators' PZTs, depending on which quadrature combination we want to analyze. To measure at $\theta_j =0$ we use the HD2-dc and HD1-dc. To lock at $\theta_j=\pi/2$ we use the lock-in amplifiers outputs \#1 (Zurich instruments, MFLI-5MHz) and \#2 (Stanford Instruments, SR844).

\subsection*{Pump light}
At optimum conditions, the single-pass SFG conversion efficiency is  $\approx$ 5.5\%/W$^2$, producing more than 900 mW of blue light when the full lasers' power is sent to the nonlinear crystal. In most cases, we run the SFG module at powers between 25\% and 70\% of the NOPO oscillation threshold (80 - 220 mW) to reduce the overall power fluctuations due to temperature and input power drifts.  We found that the long-term power fluctuations were $< 1\%$. No intensity stabilization was applied to reduce this effect. Pump power drifts are responsible for parametric gain modulations that can degrade the generated entanglement at low frequencies.\par
\subsection*{Parametric gain and oscillation threshold}
We inject an intense seed at 1064 nm ($\approx$ 100 mW) in the optical cavity via an HR mirror to check the nonlinear gain. We use this intense injected beam only to calibrate the oscillation threshold and generate coherent amplified (signal and idler) outputs to align and mode-match the local oscillators for homodyne detection.\par

\subsection*{LOs and homodyne detectors}
The coherent local oscillators are generated by filtering the laser light with mode-cleaning cavities (MCC). The cavities consist of two plane mirrors and a curved mirror in a triangular configuration, with the bandwidth $\approx$ 1 MHz. Built in monolithic aluminum blocks, they perform spatial mode filtering and eliminate beam pointing jitter. In each homodyne detector, we overlap the modes with the corresponding LOs on a 50/50 beam splitter with the visibility $\mathcal{V}_j$. The non-unitary visibilities are converted into an effective mode-matching efficiency $\eta_j^{\text{mm}} = \mathcal{V}_j^2$. The beam splitter outputs are sent to  PIN photodiodes, InGaAs-based (Fermionics, FD500N-1064) for 1064 nm, and silicon-based (Hamamatsu, S5971) for 852 nm. The estimated quantum efficiencies $\eta_1^{\mathrm{det}}$ and $\eta_2^{\mathrm{det}}$ are shown in Table \ref{tab:eff}. We calibrate the shot-noise level by blocking the output paths of the NOPO. The detectors provide more than 18 dB shot-noise clearance above the electronic noise with LO powers $\approx$ 500 $\mu W$. \par
All photodiodes are used without protection windows to reduce optical losses. Since the PIN photodiodes are based on different semiconductors, their capacitance is different, inducing a phase delay between the photocurrents. We compensate the photodiodes capacitance mismatch by carefully designing the transimpedance amplifier to improve the broadband correlation measurements.\par
 The local oscillators' optical powers are adjusted to match the shot-noise levels of the two homodyne detectors. To reduce leakage of classical noise from the local oscillators, we optimize the balanced detectors' Common Mode Rejection Ratio (CMRR). We operate the homodyne detectors with $\mathrm{CMRR}_{1064} =\mathrm{CMRR}_{852}\approx40$ dB for frequencies below 500 kHz. 
\subsection*{Homodyne angle control}
The method used to guarantee double resonance of the NOPO cavity creates spurious back-reflected fields $c_j$ that are amplified by the parametric gain and follow the same path to the homodyne detection as the entanglement. This field interferes with the local oscillators $\beta_{\text{LO},j}$ making it possible to probe the phase of the entangled beams and select which quadrature to measure.\par
The DC signal from the homodyne detectors gives $E_{\text{dc},j} \propto \beta_{\text{LO},j}c_{j}\cos{\theta_j}$.  However, the amplitude $c_{j}$ should be maintained as low as possible to reduce classical noise contamination. To measure in the audio band, we use lock beam powers  1 mW for 1064 nm and 30 $\mu$W for 852 nm. The homodyne photocurrent is amplified with a low-noise amplifier with $\approx$ 30 dB gain and $\approx$ 1 MHz bandwidth, generating an error signal that can be used to lock the phases at $\theta_j = 0$. To create an error signal to lock at $\theta_j = \pi/2$, we introduce a phase modulation at 800 kHz in the injected 1064 nm lock beam and 600 kHz in the 852 nm local oscillator (see \ref{fig:setup_detailed}). The homodyne signal is sent to lock-in amplifiers and demodulated giving $E_{\text{dm},j} \propto \beta_{\text{LO},j}c_{j}\sin{\theta_j}$. $E_{\text{dc},j}$ and $E_{\text{dm},j}$ are sent to a feedback controller that generates a signal applied to the LO piezos.\par
\subsection*{Data acquisition and processing}
We acquire the photocurrents using a low-noise 16 bits analog-to-digital converter (ADC) (Spectrum M2p5913-x4) at a sampling rate of 5 MHz. After digital conversion, the quadratures are combined with relative weights for correlation analysis. Figure \ref{fig:measurement_results_combined} was generated by averaging 1000 spectra measurements. Each spectrum is produced with a FFT done over 16000 samples.\par

\end{document}